\documentclass[traditabstract]{aa} 
\usepackage{natbib}
\bibpunct{(}{)}{;}{a}{}{,} 

\usepackage{graphicx}
\usepackage{txfonts}
\begin{document}

   \title{Is GRB 050904 at z=6.3 absorbed by dust?}


   \author{G. Stratta
          \inst{1,2}
          \and
          S. Gallerani\inst{2}
          \and
          R. Maiolino \inst{2}
          }

  \institute{
ASI Science Data Center, via Galileo Galilei, 00044, Frascati, Italy
\and
   INAF-Osservatorio Astronomico di Roma, via di Frascati 33, 00040
              Monte Porzio Catone, Italy
}
   \date{Received ....; accepted ....}

 
  \abstract
{
Dust is an important tracer of chemical enrichment in primeval galaxies and it has also important implications for their evolution. So far, at $z>$6, close to the re-ionization epoch, the presence of dust has been only firmly established in quasar host galaxies, which are rare objects associated with enormous star formation rates.
The only non-quasar host galaxy, with modest star formation rate, for which dust extinction has been tentatively detected at these early cosmic epochs, is the host of gamma ray burst GRB050904 at $z=6.3$. However, the claim of dust extinction for this GRB has been debated in the past. We suggest that the discrepant results occur primarily because most of previous studies have not simultaneously investigated the X-ray to near-IR spectral energy distribution of this GRB.  The difficulty with this burst is that the X-ray afterglow is dominated by strong flares at early times and is poorly monitored at late times. In addition, the Z band photometry, which is the most sensitive to dust extinction, has been found to be affected by strong systematics. In this paper we carefully re-analyze the Swift/XRT afterglow observations of this GRB, using extensive past studies of X-ray flare properties when computing the X-ray afterglow flux level and exploiting the recent reanalysis of the optical (UV rest frame) data of the same GRB. We extract the X-ray to optical/near-IR afterglow SED for the three epochs where the best spectral coverage is available: 0.47, 1.25, and 3.4 days after the trigger. A spectral power-law model has been fitted to the extracted SEDs. We discuss that no spectral breaks or chromatic temporal breaks are expected in the epochs of interest. To fit any UV rest-frame dust absorption, we tested the Small Magellanic Cloud (SMC) extinction curve, the mean extinction curve (MEC) found for a sample of QSO at $z>4$ and its corresponding attenuation curve, as well as a starburst attenuation curve, and the extinction curve consistent with a supernova dust origin (SN-type). The SMC extinction curve and the SN-type one provide good fit to the data at all epochs, with an average amount of dust absorption at $\lambda_{rest} = 3000~\AA$ of $A_{3000} = 0.25\pm 0.07$ mag. These results indicate that the primeval galaxy at $z = 6.3$ hosting this GRB has already enriched its ISM with dust. 
}

   \maketitle
%

\section{Introduction}

The presence of dust at high redshifts (z$>$6) is fundamental both for the formation and evolution of the stellar
populations in early galaxies, as well as for their observability.
High dust masses at z$>$6 have been detected in the host galaxy of some powerful quasars, through detection
of intense mm/submm (far-IR rest frame) dust emission 
\citep{Bertoldi2003,Beelen2006,Wang2008}.
However, these powerful quasars are very rare objects, probably lying in extreme over-density peaks in the early
universe, and hosted in extremely powerful startbursts, then forming stars at a rate higher than 1000~$M_{\odot}/yr$.
It is important to understand whether dust is also present in less extreme galaxies at z$>$6, which are more representative
of the primeval galaxy population.

The main issue in the formation of dust in the early universe is the time scale for dust production mechanisms.
In the local universe dust is mostly produced in the atmosphere of AGB stars, which require timescales ranging from
a few 100 Myr to Gyrs to enrich the ISM with significant amounts of dust. These timescales may prevent young primeval
galaxies in the early universe to have significant amounts of dust (but see also Valiante et al. 2009). Alternatively, dust production can occur in
the ejecta of SNe \citep[e.g.][]{Todini2001}, which can enrich the interstellar matter (ISM) on a very short timescale, but which can
also destroy dust through shocks \citep{Bianchi2007,Nozawa2007}.

If the dominant dust production mechanism changes over the cosmic epochs, this may result in a change in the
dust properties (e.g. traced by the extinction curve) as a function of redshift.
In the local universe and at low redshift, different systems are characterized by different extinction and attenuation curves.
The SMC extinction curve has been shown to reproduce the dust reddening of
most quasars at z$<$2.2 \citep{Richards2003}, while the effective reddening of galaxy spectra in the local Universe is generally
modeled by using the attenuation law derived by \cite{Calzetti1994}. The possibility that dust properties may evolve with redshift was first investigated by Maiolino et al. (2004) who analyzed the spectrum of the quasar SDSSJ1048+4637 at z$=$6.2, finding that the reddening in this object is described better by an SN-type extinction curve \citep{Todini2001} than through the usually adopted SMC. A more extended study has been developed by Gallerani et al. (2010), who analyzed the optical-near infrared spectra of 33 quasars with redshifts $4.0\leq z\leq 6.4$, finding that all of the reddened quasars require an extinction curve 
deviating from that of the SMC, with a tendency to flatten at $\lambda _{rest}<2000$~\AA. By performing a simultaneous fit of all reddened quasars in the sample, Gallerani et al. (2010) obtained a mean extinction curve (hereafter called MEC) at z$>$4, and the corresponding attenuation curve (hereafter called MEC$_{att}$).
The change in extinction curve at z$>$4 suggests that the dust properties at early epochs are different
with respect to dust at intermediate and low redshifts, either because of a different dust production
mechanism (e.g. dominated by SNe instead of AGB stars) or because of different dust processing
(e.g. stronger shocks destroying small grains).

High-z GRBs provide additional opportunities to study the evolution of dust properties at high redshift \citep[e.g.][]{Stratta2007,Li2008,Zafar2010,Perley2009,Perley2010}. In principle, this kind of studies may
provide more stringent constraints on dust extinction curves relative to QSOs, since the intrinsic slope of their
continuum can be inferred from the associated X-ray data, which are not affected by dust absorption.
Moreover, GRBs have been discovered at redshifts greater than the ones of the most distant quasars, e.g. GRB080913 at
z$\sim 6.7$ \citep{Greiner2009}, and GRB090423 at z$\sim 8.1$ \citep{Salvaterra2009, Tanvir2009}. Since (long)
GRBs result from the death of (early) massive stars \citep[e.g.][]{Abel2002, Schneider2002,Woosley2006}, they are expected to
be detected at even earlier epochs. For these reasons GRBs are perfect candidates for investigating both the cosmic reionization process \citep{Gallerani2008,Patel2010} and the evolution of the ISM properties through the cosmic epochs.   

GRB 050904 was a very bright burst at z$=$6.29 for which extensive multi-band afterglow follow-up has been performed \citep{Cummings2005,Tagliaferri2005,Haislip2006,Price2006,Kawai2006,Boer2006}. At
the time of its discovery, GRB 050904 was the most distant GRB ever detected, and several works have been dedicated to broad
band data analysis and modeling and to the study of the interstellar matter in its host galaxy \citep[e.g.][]{Campana2007,Gendre2007}.  
Dust extinction in GRB~050904 has been investigated by several authors, but with discordant results. By studying
only the spectral energy
distribution (SED) of the optical/near-IR data, assuming ``standard" extinction curves (i.e. those of our Milky Way
--MW-- and for the two Magellanic Clouds --SMC,LMC--),
several authors have found very low or zero values of dust extinction \citep{Kann2007,Haislip2006,Perley2010}. From wide-band spectral afterglow modeling, very low dust
extinction values were inferred by \cite{Gou2007}, assuming an MW extinction curve. \cite{Liang2009}, in contrast, find evidence of non negligible dust extinction and a best-fit to the data assuming the extinction law computed by \cite{Li2008} with evidence of the $2175~\AA$ absorption feature.  By exploiting that
X-ray emission is not affected by dust absorption, \cite{Stratta2007} have found evidence of dust extinction at a level comparable to
what was found by \cite{Liang2009}. This result was obtained both through the
simultaneous optical to X-ray afterglow SED fitting at T+0.5 day and also from the 
optical/near-IR SED fitting at later epochs.  These authors also show that an extinction curve expected from
dust produced by supernovae, such as
the one found for a z$\sim6$ QSO \citep[e.g.][]{Maiolino2004}, provides a better fit to the data than the
MW, LMC, SMC extinction curves
and the Calzetti attenuation curves. \cite{Zafar2010} have published
a new careful data reduction of the photometric data in the Z band, which is one of the photometric points most
sensitive to dust extinction and to different extinction curves.
In their work, \cite{Zafar2010} find that, at T+0.47 days, the z-band suppression is weaker than reported in the previous works and that in
general the photometry calibration in this band is affected by systematic large uncertainties. By fitting the $zYJHK_s$ SED
by assuming a power-law spectral model, leaving both the spectral index and normalization free to vary, and using
any dust extinction curve \citep[SMC and the same SN-origin extinction model used by][]{Stratta2007}, \cite{Zafar2010} find only marginal evidence of dust extinction at T+0.47 days and no evidence at later epochs. 

Given the importance of any confirmation of dust at such high redshifts, in this work we carefully revisit the
global SED analysis for this burst. In particular, contrary to the majority of past works based on optical/near-IR data SED
fitting alone, we exploit the X-ray afterglow emission of this burst to determine the spectral slope and normalization of
the intrinsic, unabsorbed continuum. The difficulty with this burst is that the early X-ray afterglow is dominated by strong flare activity. X-ray flares are thought to be a distinct component than the afterglow one, possibly linked to the initial burst production mechanism \citep[e.g.][]{Falcone2007}. 
We thus perform a new, detailed, and careful analysis of the X-ray afterglow of GRB 050904, devoting effort to estimating the X-ray flare afterglow contamination at the epochs of our interest. 
We take optical/near-IR data from the literature at T+0.47, 1.25, and 3.4 days,
where the photometric data in the Z band is taken from the accurate reanalysis recently published by \cite{Zafar2010}. 

The results of our work are reported in \S 4 where we provide evidence that, at all three analyzed epochs, the optical and X-ray afterglow data of GRB 050904 are consistent with being extincted by dust, independent of the assumed extinction law model. 


  \begin{figure*}
   \centering
   \includegraphics[width=15cm,angle=-90]{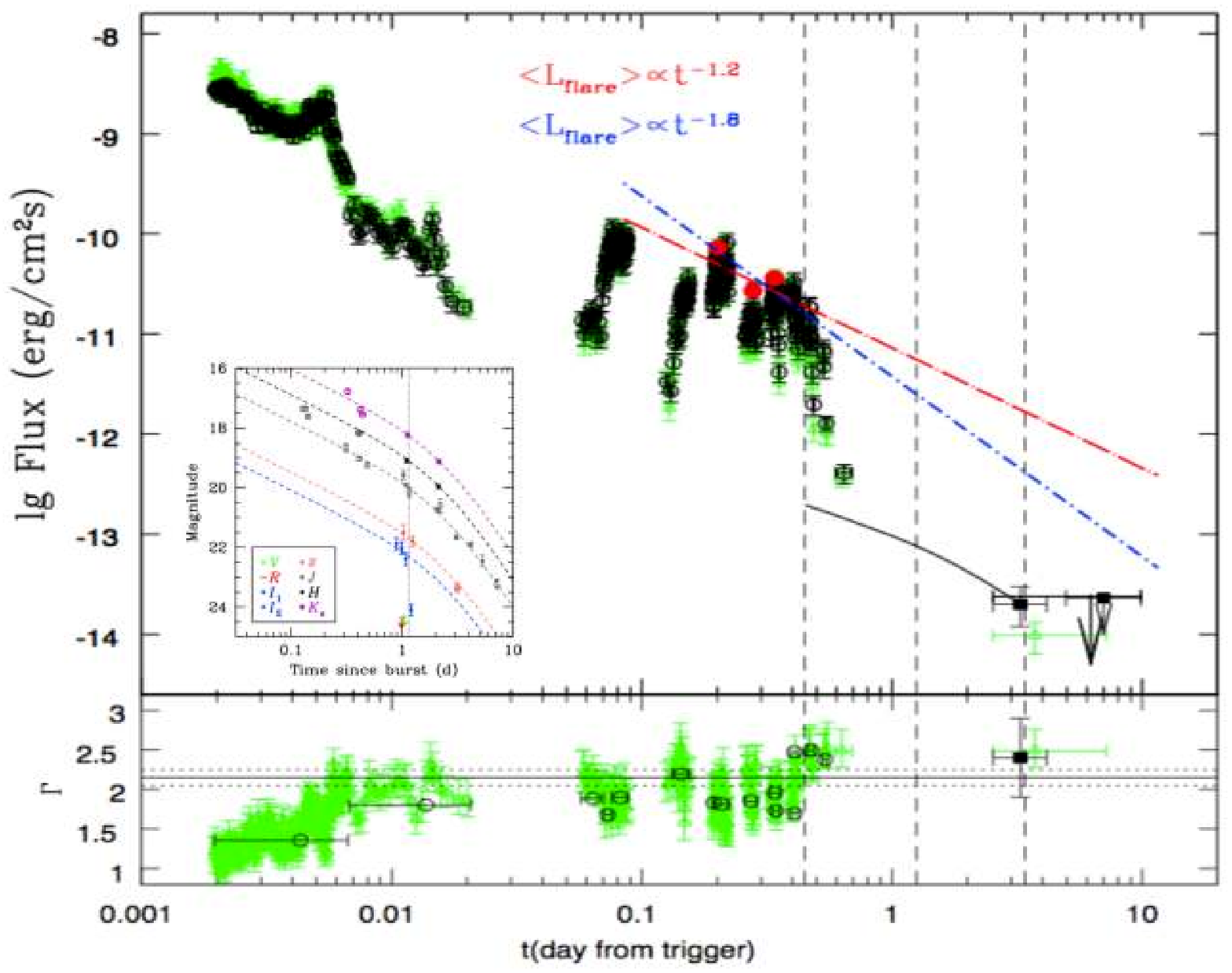}
   \caption{GRB 050904 X-ray afterglow unabsorbed flux as obtained from ``xrtgrblc'' task (open black circle) and from the ``burst
   analyzer" tool by \cite{Evans2010} (green triangles). The black solid square at about 3.254 days after the burst trigger is the
   flux estimated with the stacking method discussed in the text and used in the temporal analysis. The solid line is the best-fit smoothed broken power-law model from the optical data from Tagliaferri et al. (2005) (see inserted panel). The
   dashed vertical lines indicate the three epochs at which the largest spectral optical coverage is available from the
   literature, namely T+0.47, T+1.25, and T+3.4 days. The blue dot-short-dashed and red dot-long-dashed lines indicate the expected average flare intensity evolution with time, for two different decay rates predicted for late time flares (see \S2.1 for more details). The bottom panel shows the X-ray photon index as a function of time
   obtained by both the ``xrtgrblc" task (black circle) and the ``burst analyzer" tool (green triangles), as well as the
   estimate obtained with the stacking method at about 3.4 days after the burst (black squares).
   Horizontal dashed line indicates the expected value at the epochs of our interest where $\nu_c<\nu_X$, for a synchrotron emission from an electron population with energy spectral index of $p\sim2.1-2.5$, that is $\Gamma_{X}^{exp}=2.15\pm0.10$.}
              \label{fig:f1}%
    \end{figure*}

\begin{table*}
\begin{center}
\caption{Swift/XRT observations of GRB 050904. From left to right there are the Swift/XRT sequence (observation) identifying number, the start and end date of each sequence, and their corresponding distance in days from the burst trigger time. There are no XRT observations between 0.6904 and 2.535 days after the trigger. \label{tab:t0}}
\begin{tabular}{l c c c c }
\hline
Swift/XRT   & $T_{start}$ &   $T_{end}$ & $\Delta T_{start}$ & $\Delta T_{end}$ \\
sequence ID	& YYYY-MM-DD UT 	&  YYYY-MM-DD UT	  & day 	& day \\
\hline
00153514000 & 2005-09-04 02:01:26 & 2005-09-04  18:26:00 & 0.006738 & 0.6904 \\
00153514001 & 2005-09-06 14:41:59 & 2005-09-06  18:38:56 & 2.535  & 2.699 \\
00153514003 & 2005-09-07 00:29:33 & 2005-09-07  23:32:57 & 2.943  & 3.903 \\
00153514004 & 2005-09-08 00:48:36 & 2005-09-08  05:42:51  & 3.956 & 4.16 \\
00153514005 & 2005-09-09 00:52:08 & 2005-09-09  23:25:57  & 4.959 & 5.899 \\
00153514006 & 2005-09-10 01:05:23 & 2005-09-11 06:17:58  & 5.968 & 7.185\\
00153514007 & 2005-09-13 01:03:19 & 2005-09-13  23:59:57  & 8.966 & 9.922\\
\hline
\end{tabular}
\end{center}
\end{table*}

\section{Swift/XRT data analysis}
\label{xrtdatanalysis}

Swift/XRT data of GRB 050904 have been calibrated, filtered, and screened using the XRTDAS package included in the HEAsoft
distribution (v6.9), as described in the XRT Software User's Guide\footnote{$http://swift.gsfc.nasa.gov/docs/swift/analysis/$}. The 0.3-10 keV light curve of
this burst is flare-dominated at almost all epochs, at least up to one day after the trigger time T at Sept 5.0785, 2005 \citep{Cummings2005}. Given the flare time-varying energy spectrum \citep[e.g.][]{Falcone2007}, time-resolved spectral
information was taken into account when converting counte-rates to energy fluxes. This procedure was performed with
two methods. In the first one, we used the  ``xrtgrblc" task. This task extracts the light curve from all the XRT
observations (sequences) performed for this burst (Table \ref{tab:t0}), optimizing the source and background extraction regions and
taking pile-up into account in a time-dependent manner. The conversion of the counte-rates into flux units is performed on the basis of an  energy spectra extraction in temporal bin with width that depends from the counte-rate, and on the best-fit of each extracted spectrum
assuming a double absorbed power-law spectral model, with one absorption component fixed at the Galactic value. Final fluxes are corrected for total absorption.
In the second method, rather than extracting the spectrum in a count-rate dependent temporal region and fitting the power-law model to the data (as the ``xrtgrblc" task does), the count-rate to flux conversion is performed by assuming the power-law photon index that best reproduce the observed hardness ratio and its evolution. This method is applied in the ``burst analyzer" tool\footnote{$http://www.swift.ac.uk/burst\_analyser/$} \citep[see more details in][]{Evans2007,Evans2010}.
Comparing the light curves obtained with the two methods described above, we find overall agreement
in terms of unabsorbed fluxes and spectral slopes,
within errors (Fig.\ref{fig:f1}). 

However, at late times (t$>$T+2 days) the uncertainties on the estimated flux are quite large. 
A possible reason may be attributed to the very low flux level of the afterglow at this epoch: while the automatic time-dependent spectral analysis obtained with the ``xrtgrblc" task and through the burst analyzer tool are extremely helpful for the first, bright part of the light curve, they may be not suitable for such faint detection levels. 
To accurately estimate the flux at late times we perform an alternative data analysis 
focused on this part of the light curve.  
We progressively stack the observations starting from $T+2.535$ days  (sequence number 00153514001, see Table \ref{tab:t0}) up
to the end of the XRT observation at T+9.922 days. The stacking of the cleaned event files was performed through the
Xselect v2.4a software package, while the exposure maps were summed through the Ximage v4.5.1 software package. 
Since the source flux is fading, we find a maximum signal-to-noise ratio if we stop our stacking at $T+4.16$ days (sequence number 00153514004).  
Using the temporal range found above, we extract the source and background energy spectrum from a circular region with a 10
pixel radius and an annulus region with an inner radius of 15 pixels and an outer radius of 30 pixels, respectively.
By grouping the spectrum requiring a minimum of 5 counts per energy bin, we obtain a source energy spectrum with three energy bins. The model fitted to the obtained spectrum is an absorbed power-law model, with an absorption component fixed at the Galactic value in the direction of this burst ($4.6\times10^{20}$cm$^{-2}$, Kalberla et al. 2005). Past results agree with no evidence of additional absorption at this epoch \citep[e.g.][]{Campana2007,Gendre2007}. We find a best-fit photon index of $\Gamma=2.4\pm0.5$ (68$\%$ confidence range) and a 0.3-10 keV unabsorbed flux of $(2.0^{+1.0}_{-0.8})\times10^{-14}$ ergs cm$^{-2}$ s$^{-1}$ (68\% confidence range). 
The obtained flux has been sed at the photon weighted mean epoch, that is at $T+3.254$ days, with a start time of $T+2.535$ days and an end time of $T+4.160$ days.

The light curve we use in the following timing analysis is the one obtained from the burst analyzer tool since it provides a better temporal resolution of the spectral properties. For late time observations we use the flux estimated with our faint source dedicated analysis as explained above.

\subsection{X-ray flare analysis}
\label{xlightcurve}

To compare the spectral energy distribution (SED) of the optical afterglow with its X-ray counterpart, we need the X-ray flux level at those epochs with the best optical spectral coverage, which is at T+0.47, T+1.25 and T+3.4 days \citep[e.g.][]{Zafar2010}. At T+1.25 days no XRT data are available (Table 1). At T+0.47 days the X-ray light curve is flare-dominated, while the simultaneous afterglow optical counterpart does not show evidence of such intense flaring activity (Fig.\ref{fig:f1}). There is general consensus about considering flares to have a different origin than the afterglow \citep[e.g.][]{Falcone2007}, therefore we need to estimate the X-ray afterglow flux level excluding the flares. Near T+3.4 days there are XRT data, centered at T+3.254 days, thus requiring only a small temporal extrapolation. 

Before proceeding to the afterglow flux extrapolation at the epochs of interest, we need to know whether late-time (i.e. after $\sim T+0.5$ day) XRT data are still contaminated by flares or not.  
We know that the peak-to-continuum flux ratio $\Delta F/F$ of X-ray flares spans from $\sim1$ up to 1000 in some extreme cases (typically very early single flares): indeed, the median of $\Delta F/F$ computed over tens of GRBs and hundreds of flares \citep{Chincarini2010} is about $4$ and at late times is expected to be 1.
Therefore, even if flare activity is present at late times for GRB 050904, we expect that the XRT data at $T+3.254$ days would represent the afterglow flux level enhanced at worst by a factor of a few. However, in the following we provide some arguments that strongly support the idea that flare activity has deeply damped or ceased at late times for this burst.

Past studies have shown that the average X-ray flare peak luminosity $<L>$, corrected for the contribution of the underlying afterglow power-law component, decays with time as $<L>\propto t^{-\alpha}$, with $\alpha$ in the range 1.5-1.8 (Lazzati et al. 2008). Based on a larger sample, recently \cite{Margutti2011} have found that $<L>\propto t^{-2.7\pm0.1}$ for flares detected up to 1000 s after the burst onset (rest frame) and $<L>\propto t^{-1.2\pm0.1}$ at later times. Since at late time flares may be affected by a low signal-to-noise ratio\footnote{Indeed, \cite{Margutti2011} have shown that the evolution of the flare flux with time resembles the average detectability threshold one.}, it has been argued that their unbiased intensity decay may be steeper. For this reason, a joint fitting of early and late time flares has been performed, providing a unique decay index of $\alpha=1.8$ \citep{Bernardini2011}. 

We exploit these results to have an estimate of the expected flare intensity at late times for GRB 050904. The X-ray flux starting from about $T+0.08$ days ($\sim T+1000$ s rest frame) is completely dominated by flares so that the underlying continuum does not affect the peak flare evolution with time. In Figure 1 we plot the average flare intensity evolution, with  two possible decay rates, where the blue line marks $\alpha=1.8$ and the red one marks $\alpha=1.2$. Each power-law has been fitted to the three best-fit flare peaks found by \cite{Falcone2007} at late times (i.e. with $T+t_{peak}>1000$ s in the rest frame) for this burst. The average flare evolution, computed assuming both $\alpha=1.2$ and $\alpha=1.8$, can trace the GRB 050904 flare peak intensity reasonably well up to about $T+0.5$ days, with a preference for the shallower decay index (in agreement with past findings, see Lazzati et al. 2008, Margutti et al. 2011). If we assume that flare activity is still present at later times, the extrapolation overpredicts the observed fluxes by more than one order of magnitude. In addition, $\sim T+0.5$ day marks a clear hard-to-soft spectral transition, as shown in the bottom panel of Figure 1. Flares are spectrally harder than the underlying continuum \citep[e.g.][]{Margutti2011}: the rather sharp spectral transition simultaneous to the flare peak intensity deviations from the average expected values is consistent with a scenario where flare activity has significantly damped, or ceased, at epochs later than $T+0.5$ days. These findings enable us to consider the X-ray flux measure at $T+3.254$ days with large confidence as representative of the flare unbiased afterglow flux level. 

\subsection{X-ray light curve modeling and flux extrapolation}

To extrapolate the X-ray afterglow flux level at the epochs we are looking at, we need to assume a temporal model. Past broad band modelings of this burst have shown that data from radio to X-rays \citep[e.g.][]{Frail2006,Gou2007} are consistent with the fireball paradigms. The afterglow emission is commonly interpreted as synchrotron radiation emitted by a population of relativistic electrons \citep[e.g.][]{Meszaros1997}.  The synchrotron cooling frequency \citep[e.g.][]{Sari1998} of GRB 050904, turns out to lie at energies below the optical range at the epochs of our interest \citep[e.g.][]{Kann2007,Frail2006}, in agreement with the absence of any optical spectral variations at those epochs (e.g. Tagliaferri et al. 2005). Thus, X-rays and optical afterglow fluxes are expected to decay jointly, following the same light curve, with no spectral break between the two energy domains. Estimated values of the electron spectral index $p$ are within the range $2.1-2.5$ \citep{Frail2006,Gou2007,Zou2006,Chandra2010}, providing an expected X-ray photon index of $\Gamma_{X}^{exp}=\beta_{X}^{exp}+1=p/2+1=2.15\pm0.10$ consistent with our late time measure. 

The optical afterglow light curve of GRB 050904 is well fitted by a smoothed broken power-law model between 0.1 and 10 days after the trigger, with best-fit decay indexes $\alpha_1=0.72\pm0.10$ and $\alpha_2=2.4\pm0.4$ and a break time at $t_b=2.6\pm1.0$ days \citep{Tagliaferri2005}. \cite{Kann2007} quote $\alpha_1=0.85\pm0.08$ and similar $\alpha_2$ and $t_b$ fitting optical data from $T+0.3$ days. The temporal break is achromatic, and therefore 
it has been interpreted as evidence for a jet with opening 
angle $\theta\sim 1/\Gamma(t_b)$, where $\Gamma(t_b)$ is 
the ejecta Lorentz factor at the epoch of the temporal break \citep[e.g.][]{Sari1999}. 

We thus simply normalize the best-fit optical light curve model (from Tagliaferri et al. 2007) at the $T+3.25$ day unabsorbed flux value (see Fig.1). The X-ray flux extrapolations at $T+0.47$, $T+1.25$ days, and $T+3.4$ days, are 0.03, 0.07, and 0.001 $\mu$Jy. We associated an uncertainty of $25\%$ to the former two values and of about $50\%$ to the last value, encompassing the uncertainty of the late X-ray flux measures.

\section{Optical to X-ray spectral energy distribution}



\begin{table*}
\centering
\caption{Best-fit model parameters obtained for $\beta_{X,opt}=\beta_X~\epsilon~[0.9;1.9]$, i.e. $1\sigma$ interval of the X-ray energy spectral index value.}\label{tab:t4}
\begin{tabular}{l l l l l l}
\hline
\newline
Epoch & Extinction curve&$\beta_{X,opt}$&$A_{3000}$&$\tilde{\chi}^2_{BF}$&$P(\tilde{\chi}^2<\tilde{\chi}^2_{BF})$\\
\hline
0.47 days &{\bf SMC}  & $1.15^{+0.04}_{-0.04}$ & $0.36^{+0.01}_{-0.01}$ &{\bf 0.92} &{\bf 0.60} \\ 
		  & MEC 		  & 1.19					& 0.69				& 1.68 & 0.81 \\ 
	 	  & Calzetti 	  & 1.22					& 0.99	 			& 1.35	& 0.74 \\ 
		  & MEC$_{att}$   & 1.29					& 1.50 				& 2.37 & 0.91 \\ 
		  &{\bf SN-type}  & $1.16^{+0.09}_{-0.09}$ &$0.51^{+0.03}_{-0.03}$  &{\bf 0.02} &{\bf 0.02} \\ 
\hline
1.25 days & {\bf SMC}  & $1.11^{+0.11}_{-0.10}$ & $0.07^{+0.04}_{-0.02}$ &{\bf 0.07} &{\bf 0.02} \\ 
		  & MEC 		  & $1.12^{+0.11}_{-0.11}$	& $0.14^{+0.06}_{-0.09}$ & 0.09	& 0.04 \\
	 	  & Calzetti  	  & $1.13^{+0.11}_{-0.11}$  & $0.22^{+0.07}_{-0.17}$	& 0.08	& 0.03 \\ 	 	  
		  & MEC$_{att}$   & $1.15^{+0.11}_{-0.12}$	& $<0.43$ 			& 0.11 & 0.04 \\ 
		  &{\bf SN-type}  & $1.11^{+0.11}_{-0.10}$	& $0.09^{+0.04}_{-0.03}$ &{\bf 0.11} &{\bf 0.04} \\ 
\hline
3.4	days  &{\bf SMC}  & $1.12^{+0.09}_{-0.08}$ 	& $0.19^{+0.02}_{-0.01}$&{\bf 0.13} &{\bf 0.12} \\ 
		  & MEC 		  & $1.15^{+0.08}_{-0.04}$	& $0.38^{+0.03}_{-0.07}$				& 0.30 & 0.26 \\ 
	 	  & Calzetti	  & $1.17^{+0.08}_{-0.18}$  & $0.57^{+0.02}_{-0.15}$ 				& 0.22	& 0.20 \\ 
		  & MEC$_{att}$   & $1.21^{+0.08}_{-0.08}$	& $0.90^{+0.04}_{-0.28}$& 0.46 & 0.37 \\ 
		  &{\bf SN-type}  & $1.13^{+0.09}_{-0.09}$	& $0.27^{+0.03}_{-0.03}$&{\bf 0.04} &{\bf 0.04} \\ 
\hline
\end{tabular}
\end{table*}


\begin{table*}
\centering
\caption{Best-fit parameters obtained for $\beta_{X,opt}=\beta_X~\epsilon~[0.4;2.4]$, i.e. $2\sigma$ interval of the X-ray energy spectral index value.}\label{tab:t5}
\begin{tabular}{l l l l l l}
\hline
\newline
Epoch & Extinction curve&$\beta_{X,opt}$&$A_{3000}$&$\tilde{\chi}^2_{BF}$&$P(\tilde{\chi}^2<\tilde{\chi}^2_{BF})$\\
\hline
0.47 days &{\bf SMC} & $1.15^{+0.04}_{-0.04}$ 	& $0.36^{+0.01}_{-0.01}$ &{\bf 0.92} &{\bf 0.60} \\ 
		  & MEC 		  & 1.19					& 0.69 				& 1.68 & 0.81 \\ 
	 	  & Calzetti 	  & 1.22					& 0.10		 			& 1.35	& 0.74 \\ 
		  & MEC$_{att}$   & 1.29					& 1.50				& 2.37 & 0.91 \\ 
		  &{\bf SN-type}  & $1.16^{+0.09}_{-0.09}$	&$0.51^{+0.03}_{-0.03}$ &{\bf 0.02} &{\bf 0.02} \\ 
		  
\hline
1.25 days &{\bf SMC}  & $1.11^{+0.11}_{-0.11}$ 	& $0.07^{+0.06}_{-0.02}$&{\bf 0.07} &{\bf 0.02} \\ 
		  & MEC 		  & $1.12^{+0.11}_{-0.11}$	& $0.14^{+0.06}_{-0.09}$& 0.09	& 0.04 \\
	 	  & Calzetti  	  & $1.13^{+0.11}_{-0.11}$  & $0.22^{+0.07}_{-0.17}$	& 0.08	& 0.03 \\ 	 	  
		  & MEC$_{att}$   & $1.15^{+0.11}_{-0.12}$	& $<0.43$  			& 0.11 & 0.04 \\ 
		  &{\bf SN-type}  & $1.11^{+0.11}_{-0.10}$	& $0.09^{+0.04}_{-0.03}$&{\bf 0.11} &{\bf 0.04} \\ 
\hline
3.4	days  &{\bf SMC} & $1.12^{+0.09}_{-0.08}$ 	& $0.19^{+0.02}_{-0.01}$  &{\bf 0.13} &{\bf 0.12} \\ 
		  & MEC 		  & $1.15^{+0.08}_{-0.04}$	& $0.38^{+0.03}_{-0.07}$				& 0.30 & 0.26 \\ 
	 	  & Calzetti	  & $1.17^{+0.08}_{-0.18}$  & $0.57^{+0.02}_{-0.15}$			& 0.22	& 0.20 \\ 
		  & MEC$_{att}$   & $1.21^{+0.08}_{-0.08}$	& $0.90^{+0.04}_{-0.28}$& 0.46 & 0.37 \\ 
		  &{\bf SN-type} & $1.13^{+0.09}_{-0.09}$	& $0.27^{+0.03}_{-0.03}$&{\bf 0.04} &{\bf 0.04} \\ 
\hline
\end{tabular}
\end{table*}


\begin{table*}
\centering
\caption{Best-fit model parameters obtained for $\beta_{X,opt}=\beta_X~\epsilon~[1.05;1.25]$ (i.e. the interval of the theoretically expected X-ray photon index for an electron spectral index $p\sim2.1-2.5$ and $\nu_c<\nu_X$.)}\label{tab:t6}
\begin{tabular}{l l l l l l}
\hline
\newline
Epoch & Extinction curve&$\beta_{X,opt}$&$A_{3000}$&$\tilde{\chi}^2_{BF}$&$P(\tilde{\chi}^2<\tilde{\chi}^2_{BF})$\\
\hline
0.47 days &{\bf SMC}  & $1.15^{+0.04}_{-0.04}$ 	& $0.36^{+0.01}_{-0.01}$&{\bf 0.92} &{\bf 0.60} \\ 
		  & MEC 		  & 1.19					& 0.69				& 1.68 & 0.81 \\ 
	 	  & Calzetti 	  & 1.22					& 0.99	 			& 1.35	& 0.74 \\
		  & MEC$_{att}$   & 1.25					& 1.39				& 2.60 & 0.92 \\ 
		  &{\bf SN-type} & $1.16^{+0.09}_{-0.09}$	&$0.51^{+0.03}_{-0.03}$ &{\bf 0.02} &{\bf 0.02} \\ 		  
\hline
1.25 days &{\bf SMC} & $1.11^{+0.11}_{-0.11}$ 	& $0.09^{+0.01}_{-0.02}$&{\bf 0.07} &{\bf 0.02} \\ 
		  & MEC 		  & $1.12^{+0.11}_{-0.11}$	& $0.14^{+0.06}_{-0.01}$ 	& 0.09	& 0.04 \\
	 	  & Calzetti  	  & $1.13^{+0.09}_{-0.08}$  & $0.22^{+0.07}_{-0.08}$ & 0.08	& 0.03 \\ 
		  & MEC$_{att}$   & $1.15^{+0.11}_{-0.12}$	& $0.42^{+0.02}_{-0.31}$ & 0.11 & 0.04 \\ 
		  &{\bf SN-type} & $1.11^{+0.11}_{-0.06}$	& $0.11^{+0.02}_{-0.02}$ 	&{\bf 0.11} &{\bf 0.04} \\ 
\hline
3.4	days  &{\bf SMC}	  & $1.12^{+0.09}_{-0.08}$ 	& $0.19^{+0.02}_{-0.01}$ &{\bf 0.13} &{\bf 0.12} \\ 
		  & MEC 		  & $1.15^{+0.08}_{-0.08}$	& $0.38^{+0.03}_{-0.07}$	  & 0.30 & 0.26 \\ 
	 	  & Calzetti	  & $1.17^{+0.08}_{-0.06}$  & $0.57^{+0.02}_{-0.16}$   & 0.22	& 0.20 \\ 
		  & MEC$_{att}$   & $1.21^{+0.05}_{-0.08}$	& $0.90^{+0.06}_{-0.28}$ & 0.46 & 0.37 \\ 
		  &{\bf SN-type}  & $1.13^{+0.09}_{-0.08}$	& $0.27^{+0.03}_{-0.02}$ &{\bf 0.04} &{\bf 0.04} \\ 
\hline
\end{tabular}
\end{table*}


  \begin{figure*}
   \includegraphics[width=19cm]{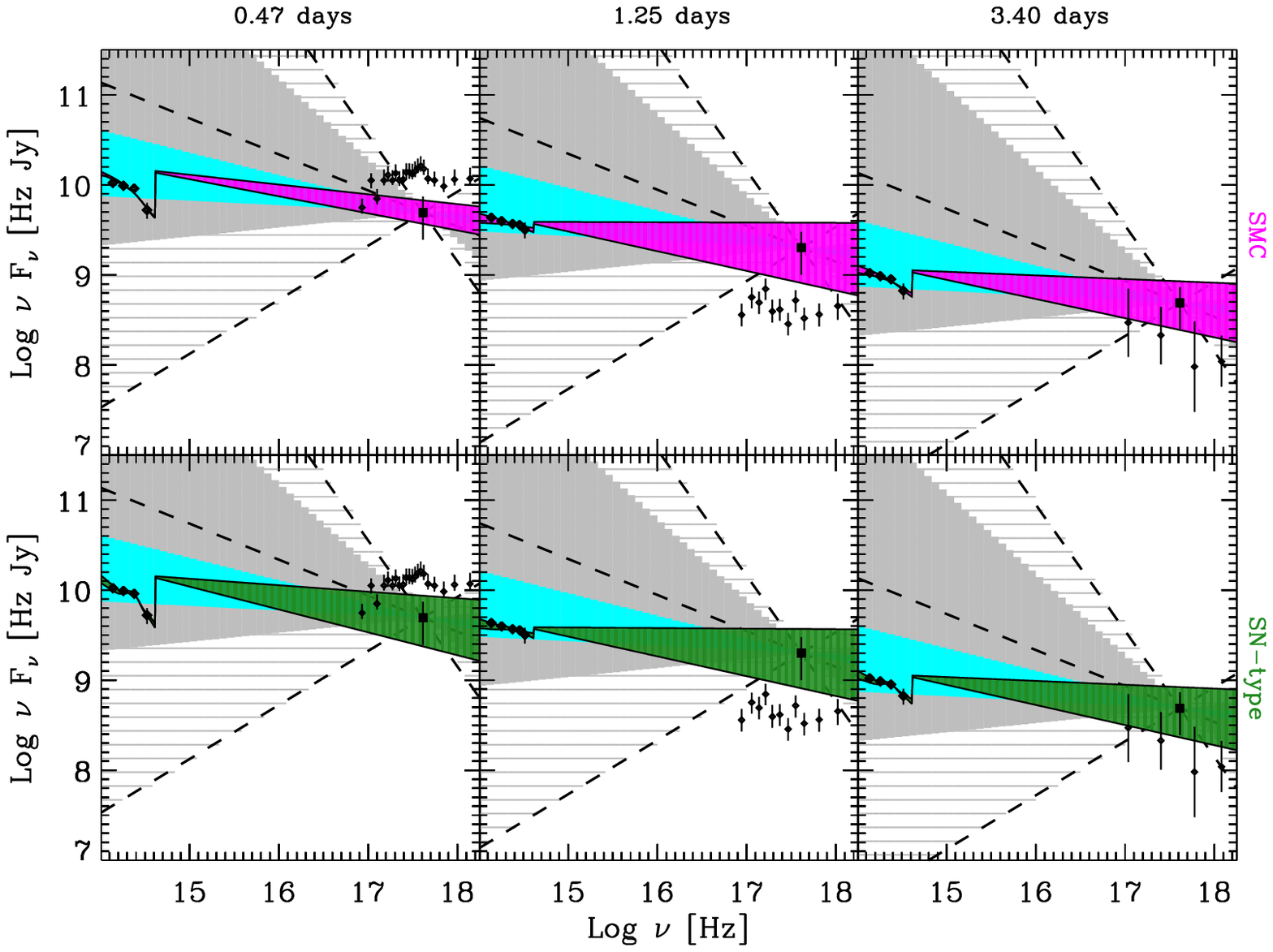}
   \caption{Optical to X-ray SED at 0.47, 1.25 and 3.4 days after the burst trigger (from left to right) and fitted using the two extinction curves that provide excellent fit at all epochs, that is, the SMC and the SN-type (from top to bottom). Squares show the X-ray flux obtained by our analysis while diamonds are the fluxes by \cite{Zafar2010}. The hatched area (delimited by dashed lines) indicates the range of intrinsic power-laws consistent with the $\pm 2\sigma$ uncertainty on $\beta_X$. The dashed line in the middle shows the power-law resulting from  the best-fit slope to the X-ray data at late epochs ($\beta_X=1.4$). The gray shaded areas show the range of intrinsic power-laws consistent with the $\pm 1\sigma$ uncertainty on $\beta_X$. The cyan shaded areas show the the power-laws expected from synchrotron emission (for an electron spectral index $p\sim2.1-2.5$ and $\nu_c<\nu_{X,opt}$). The colored areas show the optical to X-ray SED best-fit assuming a power-law model with index free to vary within the range $\beta_X\pm2\sigma$ (Table \ref{tab:t5}).}
              \label{fig:f2}
    \end{figure*}


Starting from the results obtained in the previous section, we take from the literature the optical/near-IR data corrected for Galactic absorption at T+0.47, 1.25, and 3.4 days, where the photometric data in the Z band is taken from the accurate re-analysis recently published by \cite{Zafar2010}, and we compare them with the simultaneous X-ray afterglow unabsorbed fluxes obtained in the previous section. We then fit a power-law spectral model to the data, setting the spectral index and normalization to the values obtained from X-ray afterglow. No spectral variation is expected during each epoch since $\nu_c<\nu_{X,opt}$ at these times (see \S 2.2). Given the large uncertainty affecting the X-ray photon index we have estimated at $T+3.254$ days, we conservatively performed our analysis using both the expected ($\Gamma_X^{exp}=2.15\pm0.10$, $\beta_X^{exp}=1.15\pm0.10$) and the measured ($\Gamma_X=2.4\pm0.5$, $\beta_X=1.4\pm0.5$) X-ray photon indexes.

Figure \ref{fig:f2} shows the obtained broad band SEDs at the three epochs. 
Squares give the X-ray fluxes estimated by us in the previous section, while diamonds are the values reported in \cite{Zafar2010}. We note that our X-ray fluxes deviate significantly from those reported in \cite{Zafar2010} at $T+0.47$ and $T+1.25$ days. The latter discrepancy may be due to different light curve modeling and flux extrapolation. In \cite{Zafar2010} the X-ray light curve is modeled with a smoothed broken power-law, however parameter values are not reported, hence preventing further comparative studies. 
The hatched areas (delimited by dashed lines) indicate the range of intrinsic power-laws consistent with the $\pm2\sigma$ uncertainty on $\beta_X$. The dashed line in the middle shows the power-law resulting from the best-fit slope to the X-ray data at late epochs ($\beta_X=1.4$). The gray shaded areas show the range of intrinsic power-laws consistent with the $\pm1\sigma$ uncertainty on $\beta_X$. The cyan shaded areas show the power-laws expected from synchrotron emission (for an electron spectral index $p\sim2.1-2.5$ and $\nu_c<\nu_{X,opt}$, see \S 2.2). The colored areas show the optical to X-ray SED best-fit obtained assuming a power-law model with index free to vary within the range $\beta_X\pm2\sigma$ (Tab. 3) and the extinction/attenuation curve associated with each panel. 

\begin{figure}
   \includegraphics[width=9cm]{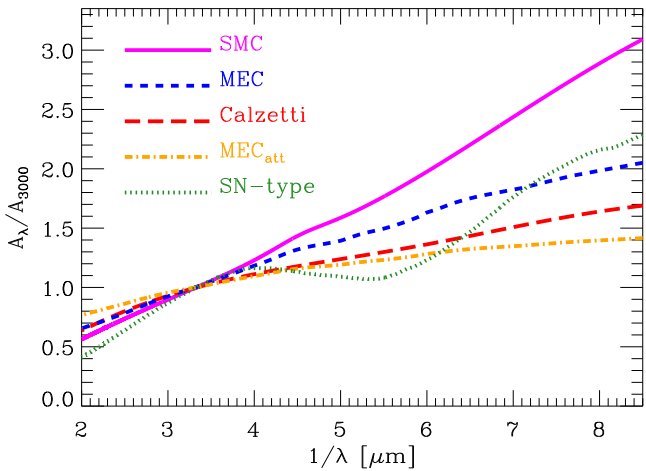}
   \caption{Extinction and attenuation curves adopted in our analysis, normalized to the extinction value at the rest frame wavelength of $\lambda_{rest}=3000$~\AA.}
              \label{fig:f3}
    \end{figure}

To quantify the need of dust extinction and reddening we have fitted the data at each epoch with different extinction and attenuation curves.
The question of whether dust reddening is better described by an extinction curve or by an attenuation law depends on the geometry of the system. 
For quasars, the simple dusty ``screen'' geometry applies, but for galaxies one has to
consider that generally dust is mixed with the emitting sources (either stars or ionized gas). 
As a consequence, attenuation curves i.e. the ratio of the observed spectrum to the intrinsic total light emitted by
the whole system before dust absorption, are likely more appropriate for galaxies. 
In the case of GRBs it is not clear which geometry is more appropriate. 
The screen case is certainly more appropriate when the size of the emitting 
optical afterglow is smaller than the distribution of the dusty medium. However, 
the dust may well be the same as produced by the progenitor before the explosion. In this case the size of the dusty medium can be comparable to the one of the expanding afterglow, and an attenuation curve may be more appropriate. As a consequence, to describe dust reddening, we consider the SMC extinction curve, the Calzetti attenuation
law\footnote{It has been shown that the Calzetti law is the result of the attenuation by a medium with an SMC-type dust \citep{Gordon1997,Inoue2005}. But see also \cite{Panuzzo2007} and \cite{Pierini2004}}, the mean extinction curve (MEC) resulting from the analysis of 33 quasars at z$\geq 4$ done by \cite{Gallerani2010}, and the corresponding attenuation
curve (MEC$_{att}$). We also test the SN-type extinction curve proposed by \cite{Todini2001}, which reproduces the dust extinction observed in a BAL QSO at z=6.2 \citep{Maiolino2004,Gallerani2010}. In Figure \ref{fig:f3}, we plot the extinction and attenuation curves adopted in our analysis, normalized to $A_{3000}$, i.e. to the extinction value at the rest frame wavelength of $\lambda_{rest}=3000$~\AA.

We report the best-fit results for each extinction/attenuation curve and at each epoch 
in Tables \ref{tab:t4}, \ref{tab:t5}, and \ref{tab:t6}, where we give the best-fit $\beta _X$ and $A_{3000}$, as well as the resulting
best-fit reduced $\tilde{\chi}^2$ and the associated probability.
Within each table we leave $\beta_X$ free to vary in different intervals:  $\beta _X~\epsilon~$[0.9-1.9] in Table \ref{tab:t4} and $\beta _X~\epsilon~$[0.4-2.4] in Table \ref{tab:t5}, which correspond to the 1$\sigma$ and 2$\sigma$ errors on the measured X-ray photon
index at late times, respectively (see \S2.1); $\beta _X~\epsilon~$[1.05-1.25] in Table \ref{tab:t6} is the range expected from synchrotron emission with $\nu_c<\nu_{X,opt}$ and $p\sim2.1-2.5$ (see \S2.2). 
The data in bold highlight the extinction/attenuation curve giving a good fit to the data ($P(\tilde{\chi}^2<\tilde{\chi}^2_{BF})<68$\%) at all epochs. 

In some cases the $\beta_X$ best-fit value coincides with the lower extreme of the interval considered. In these cases, we cannot compute the 1$\sigma$ error on $A_{3000}$ and we can only provide an interval within which this parameter is allowed to vary. Moreover, 
for some combinations of $\beta_X$ intervals and extinction curves, the best-fit values are characterized by a $P(\tilde{\chi}^2<\tilde{\chi}^2_{BF})>68$\%; in these cases, we only provide the best-fit values.

\section{Results}

Tables \ref{tab:t4} to \ref{tab:t6} show that, regardless of the interval adopted for $\beta_X$ and taking only the best-fit with $P(\tilde{\chi}^2<\tilde{\chi}^2_{BF})<68$\% into account, all epochs require dust extinction. The SMC and SN-type extinction curves provide a good fit to the data at all the three analyzed epochs (Fig.2).  

At $T+0.47$ days and at $T+1.25$ days the obtained extinction values are systematically higher and lower, respectively, than those obtained at $T+3.4$ days. Since such an evolution with time has no physical explanation, we interpret these discrepancies as the result of the systematics affecting the optical data reduction (see Zafar et al. 2010). By averaging the results obtained assuming the SMC and the SN-type extinction curve over all epochs, the amount of dust absorption at $\lambda_{rest}=3000$~\AA\ is at a level of $<A_{3000}>=0.25\pm0.07$ mag where the uncertainty takes our ignorance on the true extinction curve into account. The average best-fit optical to X-ray spectral index, independent of its initial fixed range (Tables 2, 3 and 4), is $<\beta_{X,opt}>=1.13\pm0.22$, which is nicely consistent with the range of values expected from past broad band modeling of this burst, which is $p/2$ with the electron spectral index in the range $p\sim2.1-2.5$ \citep{Frail2006,Gou2007,Zou2006,Chandra2010}.

We find that the average properties of X-ray flares \citep{Margutti2011,Bernardini2011,Chincarini2010} provide convincing indications that flare activity is strongly damped or has ceased at late times for GRB 050904. We estimate the transition from flare-dominated to afterglow-dominated X-ray flux at about $T+0.5$ day, where the expected average flare peak intensity starts to overpredicts the observed fluxes and when an abrupt hard-to-soft spectral transition is observed (Fig.1). The X-ray flux measure at T+3.254 days is more than one order of magnitude lower than the expected flare intensity, therefore likely representative of the flare-unbiased afterglow flux level. Results quoted in Tables 2,3 and 4 were obtained by extrapolating the X-ray afterglow at the three epochs where best optical/near-IR spectral coverage is available and by normalizing the best-fit temporal model obtained at optical wavelengths \citep{Tagliaferri2005} at the X-ray unabsorbed flux value measured at $T+3.254$ days (Fig.1). 

It may be argued that several GRBs show that X-ray and optical afterglow light curves do not decay with the same behavior \citep[e.g.][]{Panaitescu2006}. However, even though there is also a non negligible fraction of GRBs for which the optical and X-ray emission decay jointly \citep[e.g.][]{Oates2011}, GRB 050904 is one of the few bursts for which a broad band modeling was feasible from radio to X-rays. In particular, it has been found that data are generally consistent with fireball paradigms, with the synchrotron cooling frequency below the optical range at the epochs of interest, thus supporting our assumption \citep[e.g.][]{Frail2006}.  In addition, results for dust extinction at $T+0.47$ and $T+1.25$ days are on average consistent with the $A_{3000}$ values obtained at $T+3.4$ days, when the Swift/XRT observations are available at $T+3.254$ days, and thus flux extrapolation depends little on the assumed temporal model. At $T+3.4$ days, the extracted X-ray to optical SED shows clear evidence of dust extinction, with $A_{3000}$ in the range that goes from 0.2 to 0.9 mag, depending on the assumed dust recipe (Tab. 4). 

These results indicate, beyond a doubt, that the primeval galaxy at z=6.3 hosting this GRB has already enriched its ISM with dust. However, while the presence of dust attenuating the GRB 050904 optical afterglow at any epoch is firmly established, the type of extinction/attenuation curve is not well constrained, although we find that only the SMC and SN-type extinction curves provide a good fit to the data ($P(\tilde{\chi}^2<\tilde{\chi}^2_{BF})<68$\%) at all epochs (Fig. 2).


\section{Summary and conclusions}

We reanalyzed the afterglow of GRB 050904 at z=6.3 at those epochs where the best spectral coverage is available in the optical/near-IR range (UV rest frame), namely at 0.47, 1.25, and 3.4 days after the trigger, by fitting the simultaneous optical/near-IR and X-ray SED. In this work, we exploited the recent reanalysis done by \cite{Zafar2010} on Z-band data, which are the most sensitive to dust reddening for this burst.

We analyzed the Swift/XRT data of GRB 050904, taking special care to estimate the flare unbiased afterglow flux level at the epochs of interest and to extrapolate the flux where no data are available. A spectral power-law model was fitted to the extracted SEDs, setting the spectral index and normalization at the dust-unbiased X-ray emission values. No spectral breaks or chromatic temporal breaks between X-rays and optical wavelengths are expected at these epochs, according to past broad band modeling results for this burst \citep[e.g.][]{Frail2006,Gou2007}. Indeed, this predicts that the synchrotron cooling frequency $\nu_c$ is already below the optical range at the time of interest, thus an identical afterglow decay law at X-ray and optical wavelengths is expected. 

We investigated any presence of dust extinction in the GRB afterglow by using the SMC extinction curve and the Calzetti attenuation law. We also used the MEC at 4.0$<$z$<$6.4 inferred by \cite{Gallerani2010} from the analysis of 33 quasars, and the associated attenuation curve (MEC$_{att}$). The SN-type extinction curve, proposed by \cite{Todini2001}, which reproduces the dust extinction observed in a BAL QSO at z=6.2 \citep{Maiolino2004,Gallerani2010} was also tested.

From a simultaneous fit of the rest frame UV to X-ray SED, we find clear evidence of dust absorption at all epochs. The SMC and SN-type extinction curves provide good fits at all epochs, with an average value of dust extinction of $0.25\pm0.07$ mag at $\lambda_{rest}=3000$ ~\AA. At $T+3.4$ days where nearly simultaneous X-ray and optical data are available, thus where X-ray temporal extrapolation weakly depends on the assumed model, the extracted X-ray to optical SED shows clear evidence of dust extinction, with $A_{3000}$ in the range that goes from 0.2 to 0.9 mag, depending on the assumed dust recipe (Tab. 4). 

Our findings indicate that the primeval star-forming galaxy at $z=6.3$ hosting this GRB has already enriched its ISM with dust. The type of extinction/attenuation curve is not well constrained by the data, although we find that only the SMC and SN-type extinction curves provide good fit to the data at all epochs.

We emphasize that our results were obtained from a simultaneous optical to X-ray SED fitting, while most of the previous studies were 
performed on the optical/near-IR SED alone. Our method strongly benefits from the information from past broad band modeling and from the X-ray afterglow flux normalization level and spectral slope at 3.25 days after the trigger, when we demonstrated that the flare activity has very likely ceased.

\begin{acknowledgements} 
We thank the anonymous referee for his/her comments. We also thank R. Margutti and D. Coward for useful discussions. 
This work made use of data supplied by the UK Swift Science Data Centre at the University of Leicester.
\end{acknowledgements}

\bibliographystyle{aa} 
\bibliography{/Users/stratta/Documents/DRAFT/references} 

\end{document}